# Dynamic coupling of a finite element solver to large-scale atomistic simulations

**Mihkel Veske[1], Andreas Kyritsakis[1], Kristjan Eimre[2], Vahur Zadin[2], Alvo Aabloo[2] and Flyura Djurabekova[1]**

[1]Department of Physics and Helsinki Institute of Physics, University of Helsinki, PO Box 43 (Pietari Kalmin katu 2), 00014 Helsinki, Finland

[2]Intelligent Materials and Systems Lab, Institute of Technology, University of Tartu, Nooruse 1, 50411 Tartu, Estonia

E-mail: mihkel.veske@helsinki.fi

**Abstract**

We propose a method for efficiently coupling the finite element method with atomistic simulations, while using molecular dynamics or kinetic Monte Carlo techniques. Our method can dynamically build an optimized unstructured mesh that follows the geometry defined by atomistic data. On this mesh, different multiphysics problems can be solved to obtain distributions of physical quantities of interest, which can be fed back to the atomistic system. The simulation flow is optimized to maximize computational efficiency while maintaining good accuracy. This is achieved by providing the modules for a) optimization of the density of the generated mesh according to requirements of a specific geometry and b) efficient extension of the finite element domain without a need to extend the atomistic one. Our method is organized as an open-source C++ code. In the current implementation, an efficient Laplace equation solver for calculating the electric field distribution near a rough atomistic surface demonstrates the capability of the suggested approach.

Keywords: multiphysics, multiscale, electric field, Laplace equation, finite element method, atomistic simulation

## 1. Introduction

Achieving atomistic spatial and temporal resolution is still challenging for experimental physics and, in many cases, numerical simulations based on well-motivated physical models are the only tools which can provide interesting insight on the atomic scale. However, due to an unavoidable trade-off between computational efficiency and desired accuracy, often seemingly promising computational models turn out to be impractical.

One way to achieve high computational efficiency and numerical accuracy is to combine continuous-space calculations with atomistic simulations like classical molecular dynamics (MD) or kinetic Monte Carlo (KMC). Some such approaches [1]–[4] have shown promising results when simulating the elastoplastic evolution of nanostructures. Others [5]–[9], being especially relevant to the present work, have used such a technique to study the effects of electric field around nanostructured materials.

When a strong electric field is applied on the surface of a metal, it induces surface charge and polarization, and under certain circumstances, it triggers field emission (FE) currents with consequent electromigration effects [10]. Thus, the high electric field may significantly affect the evolution of the system and under certain conditions might cause major surface deformations [11]. For that reason, atomistic simulations that take into account the effects of electrostatic field have a wide range of applications in atom probe tomography (APT) [12], nanoelectronics [13] and space technology [14]. Moreover, atomistic modeling is a valuable tool in the investigation of vacuum arcing phenomena (vacuum breakdowns), as the fundamental mechanisms that trigger a breakdown are not entirely clear yet. The breakdown studies are relevant to the development of new-generation linear colliders like CLIC in CERN [15], vacuum interrupters [16], free electron lasers [17] and fusion devices [18].

Simulating electronic processes on material surfaces requires an accurately calculated spatial distribution of the electric field. The common method for calculating the field around any geometry is to build a mesh around the system of interest and solve the Laplace or Poisson equation on it. The solver is usually based on the finite difference method (FDM) [9], [19], finite element method (FEM) [7] or their modifications [6]. Many authors [8], [20] calculate the electric field around nanostructures without building any mesh around it. Although such mesh-free methods might be more flexible and efficient under certain conditions, they are limited in practical applications as they incorporate only the calculation of electric field.

The mesh for solving the differential equations can be either static (it does not change during the evolution of the underlying atomistic geometry) or dynamic (the mesh is adjusted with the movement of the atoms). Both can be either structured or unstructured. The main advantage of a structured mesh is its implementation simplicity, while the unstructured one provides higher tolerance to the underlying geometry. Although the generation of an unstructured dynamic mesh requires significant computational effort, it has considerable advantages over the alternatives. Since it is reconstructed at every simulation step, its shape will accurately follow the underlying geometry with the optimal density in each region. This ensures high robustness against changes in the crystallographic structure of the material, good scalability and maximum accuracy for a given computational cost.

Effects of electric field, thus far, have been introduced in atomistic simulations based on a structured or unstructured static mesh approaches. The mesh that is generated in those works either lacks accuracy in following the underlying geometry [7] or is unnecessarily dense [6], making the total computational cost unfeasible to be performed iteratively. Also, previous works are rather not universal as they typically focus on a specific type of differential equations.

The present work is the continuation of our previous attempt to include the electronic effects in atomistic simulations by solving the Laplace equation on a structured static mesh using FDM [5]. This method enabled us to investigate the behavior of Cu surface under high electric field when small-scale surface features are present [21]–[26]. However, the high computational cost and inflexible mesh limited the earlier simulations to specific crystal structures and orientations, few nm scale and very short times. To cope with the forthcoming challenges of large scale dynamic simulations, we generalized the method by combining the dynamic mesh approach with the FEM. In this way, we provide a framework for solving multiple differential equations in vacuum and material domains, to achieve enhanced computational efficiency, scalability and tolerance with respect to the crystallographic structure of studied materials. The framework also allows us to use the results in iterative atomistic simulations like MD and KMC.

To a large extent the current work is motivated by vacuum arc studies. For that reason, we demonstrate the potential of our approach by calculating electric fields around metal nanotips which are considered to cause vacuum arcing [27], [28]. The value of the electric field that is found near the surface of the metal nanotip can be used to calculate electrostatic forces acting on atoms by the field as well as Coulomb forces due to partial charging of surface atoms as demonstrated in [5] and [29]. Those forces, in turn, perturb the atomic movement [5]. Similar conditions, i.e. presence of high electric fields around metal nanotips also appear in FE [30]–[32] and APT [6], [8], [12] studies, where our approach of combining atomistic and continuum calculations can be very useful.

## 2. Methodology

### 2.1. Overview

The main objective of the current project is to provide a tool for calculating the effects of electric field on atomistic systems for up to $10^7$ atoms with a reasonable computational effort. For that purpose, we provide an open-source C++ code that contains the modules which enable to:
- import atomistic coordinates of a nanostructure from the atomistic simulation;
- dynamically generate an unstructured mesh around the imported structure;
- solve the differential equations of interest on the mesh;
- return the solution to the atomistic simulation.

By using FEM for solving the differential equations, we can optimize the mesh density in various parts of the simulation domain. In regions of high interest, where the solution changes rapidly, the mesh can be made denser and in regions with small solution gradient and lower interest the mesh could be coarser. However, generating a mesh with appropriate density is rather obligatory as performing the calculation on a poorly optimized mesh is impractical in terms of computational cost. In our simulations, we are mostly interested in the processes that take place on the surface of the material. For that reason, the mesh we generate to follow the surface geometry is dense near the surface, becoming gradually coarser away from it. However, to meet the needs of a wider audience, such an optimization scheme can be overridden by the user.

In the following paragraphs we describe the methodology for performing the tasks listed above. Appendix A summarizes the different simulation stages with a flowchart.

## 2.2. Surface extraction

We classify the atoms of the whole material as surface, bulk and clustered or evaporated atoms. The latter ones are often present in high electric field simulations, where detachment of a part of the nanotip may happen due to field-assisted evaporation [33]. To handle such systems, we perform cluster analysis on the input atoms and separate the clustered and evaporated atoms from the rest of the material. This analysis is done by using the DBSCAN (Density-Based Spatial Clustering of Applications with Noise) algorithm [34], which allows for efficient grouping of closely packed atoms without imposing any restrictions to the system geometry.

To distinguish the surface atoms from the bulk, we use a two-step procedure. In the first step, we use a computationally efficient coordination analysis – atoms with low coordination (small number of nearest neighbors (NN)) are classified as possible surface atoms, while highly coordinated atoms are recognized as bulk. In the second step, we build a Voronoi tesselation around the atoms with low coordination number. Atoms whose Voronoi cell has at least one facet exposed to the vacuum are considered to reside on the surface, while the others are located in the bulk (see figure 1). The evaporated atoms, for which all the Voronoi facets would be exposed to the vacuum, have already been identified by the cluster analysis.

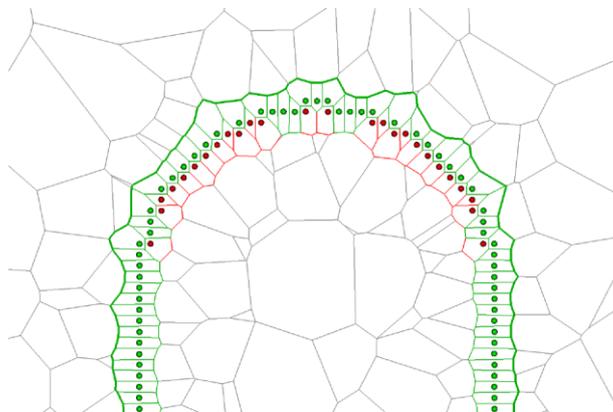

Figure 1. Slice of the Voronoi cells generated around the atoms with low coordination. The cell facets can be used to separate the surface atoms (green) from the bulk (red).

It is important to mention that such a two-step extraction is needed only in systems where the atoms are not strictly bound to the rigid crystal lattice. If the atoms do not move far away from their sites in a regular lattice, surface extraction by coordination analysis is sufficient and the computationally less efficient Voronoi cleaner can be skipped.

## 2.3. *Surface coarsening*

In general, there are two ways to generate a coarsened mesh – the top-down and the bottom-up methods. In the top-down technique, a dense mesh is generated first and then specific algorithms are used to decrease the density of elements in regions of low interest. Such an approach is useful when most of the resulting mesh is supposed to be dense and the coarsening needs to be done in a small region. The bottom-up approach, however, starts with a coarse mesh and gradually refines it until the desired quality criteria are met. The latter method turns out to be more effective if – as in our case – most of the resulting mesh is supposed to be coarse.

The total mesh generation time can be significantly reduced by making the initial, not yet refined, mesh as close as possible to the desired final one. To achieve this, it is necessary to use appropriate mesh generators, i.e. points that follow the material surface and the simulation domain boundaries and will be the nodes of the initial mesh. The generators can be obtained by designing a function which selects them among the surface atoms. Such a function should ensure that the resulting generators have the desired density in different material regions. Therefore, for every surface atom with coordinates $r$, the function should determine a clearance radius $R_{cut}$ that can be used to remove neighboring atoms which are too close (see figure 2). For instance, in systems where a cylindrical nanotip is covered with a hemisphere, we have obtained good results by using a formula as follows:

$$R_{cut}(\boldsymbol{r}) = \begin{cases} c_1 \lambda/4, & |\boldsymbol{r} - \boldsymbol{r}_{apex}| \leq R, & \text{I} \\ c_2 \lambda/4, & |\boldsymbol{r} - \boldsymbol{r}_{apex}| > R \wedge |\boldsymbol{r} - \boldsymbol{r_0}|_{x,y} \leq R, & \text{II} \\ c_3 0.1\lambda\sqrt{|\boldsymbol{r} - \boldsymbol{r_0}| - R} + c_2 \lambda/4 & otherwise, & \text{III} \end{cases} \quad (1)$$

where $R$ is the radius of the cylinder and hemisphere, $r_{apex}$ is the center of the nanotip apex, $r_0$ is the center of the nanotip-substrate junction and $\lambda$ is a characteristic distance between NN atoms. In crystalline systems $\lambda$ can be equalized to the crystal lattice constant. Parameters $c_1$, $c_2$ and $c_3$ are the integer coarsening factors that define the density of the mesh around the nanotip apex (region I in equation (1) and figure 2), nanotip lateral facets (region II) and substrate surface (region III), respectively. In general, it is necessary to specify a unique set of $c_i$ factors for each simulation geometry to meet the compromise between computational cost and solution accuracy.

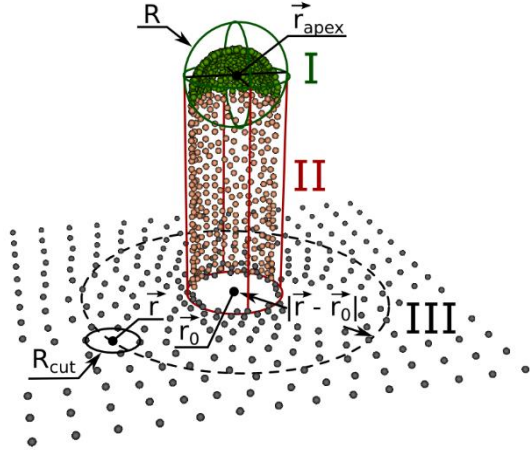

Figure 2. Principle for coarsening the nanotip surface. $R_{cut}$ – clearance radius around an atom in a position $r$, $R$ – radius of the cylinder and hemisphere, $r_{apex}$ – center of the nanotip apex, $r_0$ – center of the nanotip-substrate junction. The coloring of the nodes along with the Roman numerals designate the coarsening regions in equation (1).

It is important to mention that such a coarsening scheme can be used not only for single nanotips, but also for any geometry where the region of interest fits in a cylindrical shape. For instance, the algorithm works equally well for extrusive and intrusive formation(s) on surfaces. The geometries, whose region of interest does not fit into a cylinder or consist of several regions of interest separated by the regions of low priority, demand a customized approach. Such systems are, for instance, ridges and sparse set of nanostructures. Those geometries can be coarsened externally and imported to the simulation without a need of modifications elsewhere in the code. If such a need appears frequently, further development of the proposed algorithm can be suggested.

### 2.4. *Mesh generation*

A common way to generate a 3D FEM mesh is to build it out of tetrahedra or hexahedra. The direct generation of a tetrahedral mesh is technically much easier than that of a hexahedral one. At the same time, the basic FEM theory shows [35] that the hexahedral mesh has many advantages over the tetrahedral one, which make it more accurate and computationally efficient. For instance, the shape functions of linear hexahedral elements allow calculating a non-constant gradient for the solution inside the element – a property that is missing in linear tetrahedral elements.

Therefore, we use a hexahedral mesh that is generated in two steps. First, we generate a coarse tetrahedral mesh of suitable quality and smoothen it. Then, we split each tetrahedron into four hexahedra by appending an additional node in the centroid of each tetrahedron, triangle and line in the mesh (see the inset of figure 3). This way, we increase the computational efficiency of the FEM solver and obtain higher spatial resolution for the solution. Such mesh topology also allows us to optimize the local solution extraction process (see section 2.6).

The first-step tetrahedral mesh is generated by defining the surface atoms as described previously, define the size of the simulation box and perform a Delaunay tetrahedrization by using the open-source software package Tetgen [36]. As a result, we create tetrahedra that fill the simulation domain and pass through the generators. The next step is to check for the quality (minimum edge – outer radius ratio) and maximum volume of the resulting tetrahedra. The mesh is refined iteratively until all tetrahedra meet the specified quality and volume criteria. The details about Delaunay tetrahedrization can be found elsewhere [6], [35], [36].

The Delaunay tetrahedrization creates a mesh that fills both the vacuum and material domains. That union mesh must be separated into two because the vacuum and material domains are often handled separately. For instance, in electric field calculations only the vacuum mesh is needed, while the material domain mesh can be used to simulate processes inside the material.

In order to separate the mesh elements we use the DBSCAN algorithm [34] to create three clusters of nodes. The first cluster consists of nodes located right at the boundary between vacuum and material, the second comprises the nodes located only in vacuum and the third consists of the nodes located inside of the material. Knowing the nodal distribution allows separating also the tetrahedra. A tetrahedron will be assigned to the vacuum domain if it has at least one node in the vacuum, otherwise it shall be assigned to the material domain. This algorithm gives good results if the surface coarsening factors in equation (1) are small enough, i.e the surface nodes are close to each other. Too heavy surface coarsening results in tetrahedra that have nodes both in the vacuum and material domains and create major mesh distortions near the surface.

The mesh generated by this procedure cannot be immediately used in FEM as its surface is too rough. Apart from the atomistic roughness, there is also a high frequency noise because of the material surface is not mathematically uniquely defined. Excessive surface roughness must be removed to avoid major distortions in the FEM solution. For this purpose, many surface smoothing tools of different computational efficiency and ability to preserve sharp features in the original undistorted mesh were proposed over years [37]–[40]. A widely used algorithm providing a good compromise in the above-mentioned properties is the Taubin $\lambda/\mu$ scheme with equal weights [40]. The main advantages of the Taubin scheme are its linear spatio-temporal complexity and its ability to smoothen the surface without shrinking it. The Taubin method acts on the surface as a low pass filter by utilizing signal processing ideas. The smoothing is performed by iterating alternately the steps

$$\begin{aligned}\boldsymbol{r}'_i &= \boldsymbol{r}_i + \lambda \Delta \boldsymbol{r}_i \\ \boldsymbol{r}'_i &= \boldsymbol{r}_i + \mu \Delta \boldsymbol{r}_i\end{aligned} \quad (2)$$

where

$$\begin{aligned}\Delta \boldsymbol{r}_i &= \sum_{j \in i} \omega_{ij}(\boldsymbol{r}_j - \boldsymbol{r}_i), \\ \omega_{ij} &= \frac{|\boldsymbol{r}_j - \boldsymbol{r}_i|^{-1}}{\sum_{h \in i}|\boldsymbol{r}_h - \boldsymbol{r}_i|^{-1}}.\end{aligned} \quad (3)$$

The smoothing intensity during single iteration is controlled by the values $\lambda$ and $\mu$ that need to satisfy the constraints specified in [40]. In our test simulations, we used the values recommended by the author, namely $\lambda = 0.6307$ and $\mu = -0.6732$. The tests showed that such $\lambda$ and $\mu$ required only three iterations of equation (2) to remove most of the high frequency noise while keeping the distortions in the original geometry on an acceptable level.

The computational cost of the mesh generation could be reduced by using the tetrahedral mesh from the previous iteration as an initial guess. In the current implementation of the code, however, such a feature is not yet present and the mesh is either to be fully reused (see section 2.8.2) or to be built from the very beginning. Further optimization of the mesh generation will be a matter of future work.

### 2.5. *Calculation of the electric field*

To calculate the electric field around the nanostructure, we solve the Laplace equation on the previously described unstructured mesh and obtain the electrostatic potential $\Phi(r)$ (to minimize the computational complexity, we assume a negligible volume charge density in the vacuum):

$$\Delta \Phi = 0. \tag{4}$$

The solution of the equation (4) is the basis to calculate the electric field $\boldsymbol{E}$:

$$\boldsymbol{E} = -\boldsymbol{\nabla} \Phi. \tag{5}$$

Since we are simulating metals, we apply a Dirichlet boundary condition (BC) on the surface to obtain a constant potential there:

$$\Phi|_{surface} = 0. \tag{6}$$

A common practice in FE and breakdown studies is to assume that the anode-cathode distance significantly exceeds the linear dimensions of nanostructures. For that reason, we apply a Neumann BC on top of the simulation box to obtain a uniform long-range electric field

$$-\boldsymbol{\nabla}\Phi|_{top} = \hat{\boldsymbol{z}} E_0, \tag{7}$$

where $E_0$ is the applied field. To apply periodic BC on the lateral directions, we assume zero electric flux between the mirror images of the system, i.e

$$(\mathbf{n} \cdot \boldsymbol{\nabla}\Phi)|_{perimeter} = 0, \tag{8}$$

where $\boldsymbol{n}$ is the surface normal vector. Finally, once the mesh is generated and the BCs have been set, the Laplace equation is solved using the open-source library Deal.II [41].

The framework of the code allows adding or changing the physics without affecting the rest of the simulation flow. Such flexibility is granted by using the Deal.II library that allows solving multiple differential equations with appropriate BC-s on the same mesh. Nevertheless, one must bear in mind that the set of equations and BC-s has direct impact on the total calculation time. For instance, simulating non-conducting materials requires replacing the Dirichlet BC and calculating the charge distribution near the surface in a self-consistent manner. This is often done by means of *ab-initio* methods that involve high computational cost and are therefore beyond the scope of the current work.

### 2.6. *Local solution extraction*

To achieve acceptable computational efficiency, we use only linear hexahedral elements while assembling FEM system matrix for solving equations (4)-(8). However, the mesh is built in a way that each hexahedron has one-to-one correspondence to a tetrahedron. In a local solution extraction phase this property allows us to perform either linear or quadratic tetrahedral interpolation instead of mere linear hexahedral one. In addition to making the result smoother, the quadratic tetrahedral interpolation reduces the total interpolation time compared to a linear hexahedral one, as the number of elements that could surround the point of interest reduces by a factor of four. At the same time, we do not lose much in accuracy as the quadratic tetrahedral interpolation incorporates 10 out of 15 solution points for each tetrahedron (see inset of figure 3). Furthermore, in non-coarsened regions the characteristic tetrahedron edge length is equal to the distance between the NN atoms, giving sufficiently accurate solution around the surface atoms, even with a linear tetrahedral interpolation.

By having 10-noded tetrahedra we can choose between linear and quadratic interpolation, depending on the accuracy and computational cost requirements of the specific study case. The tetrahedral interpolation in our implementation consists of the following steps:
   a) calculate the barycentric coordinates (BCC-s) for the point of interest,
   b) use BCC-s to find the tetrahedron that surrounds the point,
   c) use BCC-s to define the shape functions and interpolate.

Although there are other ways to perform interpolation [42], the usage of BCC-s makes the computation very efficient and gives a tool for measuring the distance of any point from a triangular surface. The calculation of tetrahedral BCC-s is well standardized and the details of it can be found elsewhere [43]. Denoting by $m_{ijk}$ the $k$-th BCC of a point $i$ with respect to the tetrahedron $j$, the point $i$ is surrounded by the tetrahedron $j$, if and only if

$$m_{ijk} \geq 0 \ \forall \ k = 1,2,3,4. \tag{9}$$

Knowing the solution $\Psi_{jk}$ ($\Psi$ is a calculated quantity, e.g electric field, potential etc) on the node $k$ of the tetrahedron $j$, the interpolation at the point of interest can be obtained as

$$\Psi_i = \sum_k \phi_{ijk} \Psi_{jk}, \text{with} \sum_k \phi_{ijk} = 1 \ \forall \ i,j. \tag{10}$$

In a linear interpolation case the shape function $\phi_{ijk}$ can be equalized to the BCC [35],

$$\phi_{ijk} = m_{ijk}. \tag{11}$$

Ordering the nodes of 10-noded tetrahedron as shown in the inset of figure 3 and omitting for clarity the indices $i$ and $j$, the quadratic shape functions can be expressed as [35]

$$\phi_k = m_k(2m_k - 1), \quad k = 1,2,3,4,$$
$$\phi_5 = 4m_1m_2, \quad \phi_6 = 4m_2m_3, \quad \phi_7 = 4m_3m_1, \quad (12)$$
$$\phi_8 = 4m_1m_4, \quad \phi_9 = 4m_2m_4, \quad \phi_{10} = 4m_3m_4.$$

Note that in both cases the shape functions are properly normalized, as the BCC-s are normalized by definition, i.e $\sum_k m_{ijk} = 1 \; \forall \; i,j$.

A significant part of the computational cost of such an interpolation method consists of finding the tetrahedron that surrounds the interpolation point. The cost can be reduced by increasing the data locality, i.e by sorting the interpolation points in a way that every next point is located close to the previous one. This way every subsequent point is located in the same tetrahedron as the previous one or inside one of its neighboring tetrahedra. We achieve such a spatial ordering by sorting the points of interest along a 3D Hilbert curve [44].

### 2.7. Smoothing the results

Due to the practical need to minimize the computation time, we use only linear hexahedral elements in the FEM solver. Although the trilinear shape functions that are associated with them allow calculating a non-constant solution gradient inside the element, the gradient is discontinuous on the hexahedral faces. Such a shortcoming introduces a numerical error in the calculated electric field as different elements result in slightly different gradient of the electrostatic potential for the same node.

Another factor that has a strong impact on the solution accuracy is the quality of the mesh. The higher the symmetry of the elements and the smoother the mesh, the more accurate results can be obtained. In continuous geometries, there is no theoretical limit to refining the mesh – and thus improving the solution accuracy – while following the underlying geometry. In atomistic simulations, however, the physical limit of the accuracy is drawn by the discreteness of the atomistic system. As the surface is inherently rough on an atomic scale, the same roughness will appear in the mesh generated above it. Moreover, the atom-level roughness is accompanied by the remnants of a high frequency noise appearing during the mesh generation. In our simulations, we cannot filter out such a leftover noise completely, because heavy smoothing would cause major distortions in the atomistic geometry and would therefore significantly distort the solution.

Nevertheless, linear elements with asperities at the surface may lead to artificially enhanced electric fields on some surface nodes. A common way of avoiding such distortions is to fillet the sharp corners by defining higher order isoparametric elements and use more advanced mesh smoothing algorithms. This will increase the reliability of the solution, but also significantly reduce the computational scaling, hence, not acceptable for the purpose of the current work. To meet a compromise between solution accuracy and computational cost, we develop a smoothing algorithm that allows for increase of the signal-to-noise ratio and for elimination of the "spikes" from the electric field distribution.

The algorithm takes advantage of the fact that the largest asperities with potentially the most inaccurate solution are always located on the nodes of the tetrahedra. The rest of the nodes are on the centroids of the tetrahedra, triangles or lines and are therefore always guaranteed to have flat neighborhood at least in 1 dimension. Thus, the algorithm replaces the electric field in the tetrahedral nodes with the weighted average field on its surrounding hexahedral nodes (see figure 3). After the averaging, the electric field in the $i$-th tetrahedral node will be

$$\boldsymbol{E}_i = \frac{\sum_{j \neq i} w_{ij} \boldsymbol{E}_j}{\sum_{j \neq i} w_{ij}}, \qquad w_{ij} = \exp\left(-\frac{|\boldsymbol{r}_i - \boldsymbol{r}_j|}{L}\right), \tag{13}$$

where $\boldsymbol{E}_j$ is the electric field in the $j$-th node of all the hexahedra that contain the $i$-th node, $w_{ij}$ is the statistical weight that is a function of a distance between a tetrahedral and a hexahedral node and $L$ is the length of the longest edge in the mesh. By using such a weight function, we ensure that the electric field is distributed fairly – hexahedral nodes close to the tetrahedral one have significantly higher weight than the ones which are farther away.

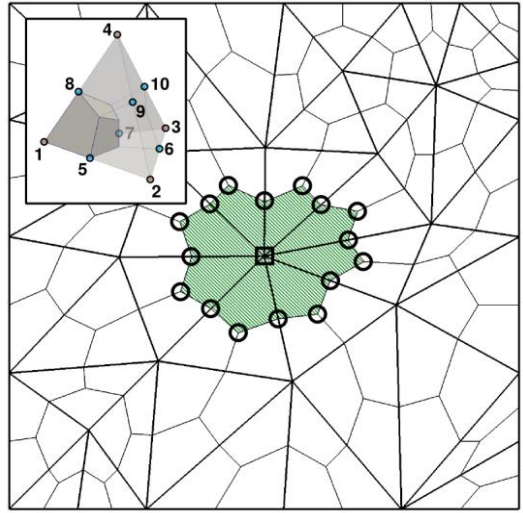

Figure 3. 2D illustration of averaging of the electric field in the nodes of the tetrahedra. Circles show the hexahedral nodes that contribute to the weighted average solution in the node of a tetrahedron, which is marked by a square. Inset: the splitting of a tetrahedron into four hexahedra. The indices correspond to the nodal ordering in 4- and 10-noded tetrahedra.

### 2.8. *Special optimization features*

Our code is designed to be combined with atomistic simulations, such as MD or KMC. To increase the overall efficiency, several features were added to the code to reduce the CPU and memory consumption of the multiscale simulations.

### 2.8.1. *Extending the simulation domain*

In atomistic simulations, the phenomena of interest are often well localized in a certain region, while the rest of the simulation domain is present to eliminate undesired effects appearing due to the periodic boundary conditions. Many such effects can be reduced by increasing the size of the simulation domain. Increasing the system size by adding extra atoms requires a significant increase of computational resources.

The usage of the FEM, however, allows building an extended simulation domain more efficiently. As the current approach was developed for simulating the processes on or near the surfaces, we extend the system by extending the surface, as demonstrated in figure 4. The shape of the extended surface can be defined either by providing an analytical formula that describes it, or by directly inputting the location of the additional nodes from a pre-built file.

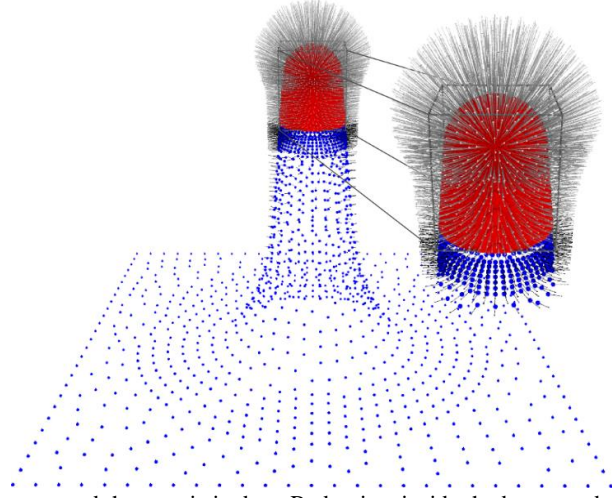

Figure 4. Extending the surface around the atomistic data. Red points inside the box – nodes from the atomistic simulation; blue points outside the box – nodes of the extended surface. Arrows around the nanotip indicate the magnitude and direction of local electric field. The arrows have different colors in MD and extended region for visual purposes.

*2.8.2. Reusing the solution*

Typically, atoms move within a few percent of an angstrom during one MD timestep. As the expected change in the system geometry is rather small, it is reasonable to expect that due to the stability of the Laplace equation [45], the change in the solution will be insignificant. Noting that due to numerical errors the calculated electric field always fluctuates, it appears that the field might be reused in several MD iterations, if the change in the system geometry is small enough.

To use such an approximation in practice, we estimate the change in the system geometry by calculating the root-mean-square distance (RMSD) the atoms have moved since the last full iteration,

$$\text{RMSD} = \sqrt{\frac{1}{N} \sum_{i \in N} \left| \boldsymbol{r}_i - \boldsymbol{r}_i^{ref} \right|^2}. \tag{14}$$

If the RMSD is below the threshold, the field in the updated location will be interpolated in the previous solution space. In case the RMSD exceeds the threshold, the whole solution will be recalculated and the reference atom coordinates reset. The advantage of the RMS value over the mean value is its higher sensitivity to major local geometry changes, as the longer displacements contribute more to the total sum than the shorter ones.

## 3. Results

The described algorithms are organized as an open-source software package called FEMOCS that can be freely downloaded from [46]. The proposed code can be used either as a standalone application or as a library. For the latter case we implemented C, C++ and Fortran interfaces that enable the usage of FEMOCS with only minor modifications in the main code. In the standalone mode, the code can read atomistic coordinates from a file and run the mesh generator and differential equation solver for these data. There is also an option to omit the atomistic section, import the mesh from a file and solve the differential equations on it.

In the following sections, we demonstrate the accuracy, speed and robustness of the code by running it on several test cases.

### 3.1. Validation of the model

We validate our code by calculating the electric field around a hemisphere on a flat planar substrate (figure 5), as this is a geometry for which the Laplace equation has a well-established analytical solution. The appendix B contains more details on the formulas, along with an estimation of the box-size-related systematic errors.

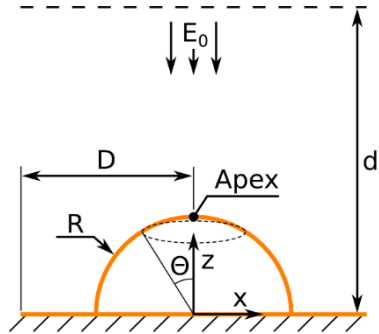

Figure 5. The geometry used to numerically calculate the field distribution. On a rectangular substrate with a thickness of 1nm and a width of $2D = 10R$, there is a hemispherical nanotip with radius $R$ that is exposed to the vacuum and to the electric field. The uniform electric field $E_0$ is applied as a boundary condition at a distance $d$ from the surface of the substrate. Polar angle $\theta$ is used during the error analysis. The system is periodic in lateral directions.

To demonstrate the numerical stability of the code, we first test it on a pseudo-atomistic hemispherical system with high rotational symmetry. We construct the hemisphere by placing the nodes symmetrically to the $z$-axis and by ensuring the characteristic distance $\lambda$ between the NN nodes. This way we can generate a smooth mesh which does not introduce "spikes" into the electric field distribution. Thus, it is possible to estimate the impact of the field post-processor to the solution accuracy and demonstrate the convergence of the numerical solution to the analytical one.

Furthermore, to test the method on a more realistic atomistic system, we replace the symmetric smooth surface with a faceted atomistic one. For this we cut the hemisphere and substrate out from a ⟨100⟩ face-centered cubic (FCC) single crystal with a lattice constant of 3.61 angstroms (Cu).

The solution accuracy depends on the mesh density which can be varied by specifying the maximum tetrahedron volume or by altering the coarsening factors of equation (1) as

$$c_1 = c_2 = c_3 = c = 0, 1, 2, \ldots \quad (15)$$

For the smooth surface system, the density can also be controlled by varying the $\lambda$ parameter. Note that the latter scheme cannot be used while coupling the code with atomistic simulations where the characteristic distance between the NN atoms cannot be arbitrarily changed.

### 3.1.1. Solution accuracy in vacuum

The comparison of the analytical solution in space with the one obtained from our code for the atomistic system is shown in figure 6. The graphs show the isolines of the electric field as calculated analytically (a, c) and numerically for two different mesh spatial densities (b, d). As can be seen, the calculated solution progressively deviates from the analytical one with increasing distance from the surface. This error can be reduced by decreasing the maximum allowed volume of the elements. Nevertheless, the solution accuracy on the surface is mainly determined by the size and quality of the mesh in its vicinity. Therefore, although a denser spatial mesh may be required in some of the potential applications of the developed code, it is mostly unnecessary when simulating surface phenomena.

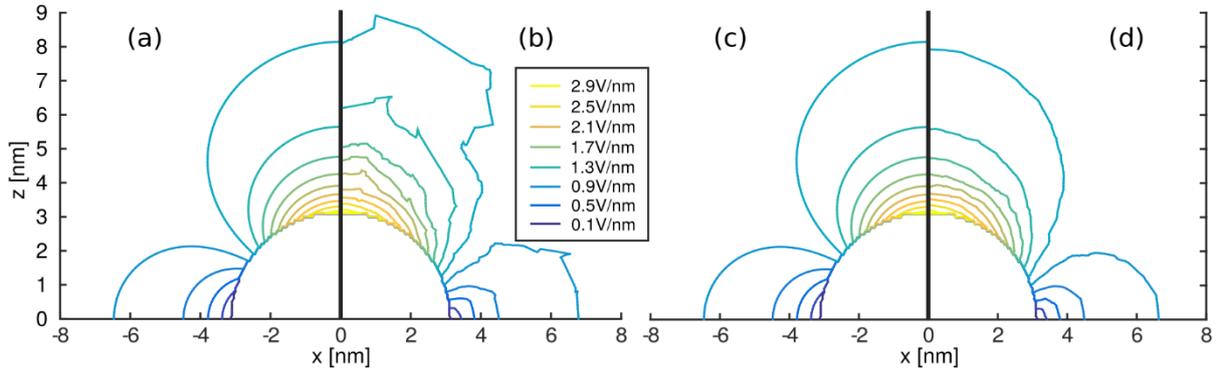

Figure 6. Electric field distribution around a hemisphere with a radius of 3 nm located on a flat surface.
(a) & (c) – analytical result, (b) – numerical result on a coarse mesh with maximum element volume $V_{max}$ = 5nm$^3$, (d) – numerical result on a dense mesh with $V_{max}$ = 0.05nm$^3$. In all cases the long-range electric field $E_0$ = 1 V/nm.

### 3.1.2. Solution accuracy on smooth surface

To demonstrate the solution accuracy near smooth surface, we denote the magnitudes of numerical and analytical electric fields as $E$ and $E_a$, respectively, and define the field error-value

$$\epsilon = \frac{E - E_a}{E_a}. \quad (16)$$

We use equation (16) to show the solution fluctuation that arises due to the numerical errors in the FEM. For this we measure the error (16) for all the tetrahedral nodes that lie on the hemispherical surface at a polar angle $\theta$ (a region marked with a dashed line in figure 5). Figure 7 shows the mean value of $\epsilon$ together with error bars that correspond to its standard error within 95% confidence level.

The graph confirms the general trend that coarser mesh results in less accurate field. It also shows that the error does not change significantly along the hemisphere. The same applies to the fluctuation amplitude of the error – on a coarse mesh the error fluctuates up to 3% while in a dense one the fluctuation is less than 1%. Furthermore, the mean error $\bar{\epsilon}$ is close to the relative distance between the NN nodes. For example, on a dense mesh with $\lambda/R = 1.7\%$ the numerical field deviates by $(1 \pm 1)\%$ from the analytical value, while on a coarse mesh with $\lambda/R = 8.3\%$ that deviation is $(7 \pm 2)\%$.

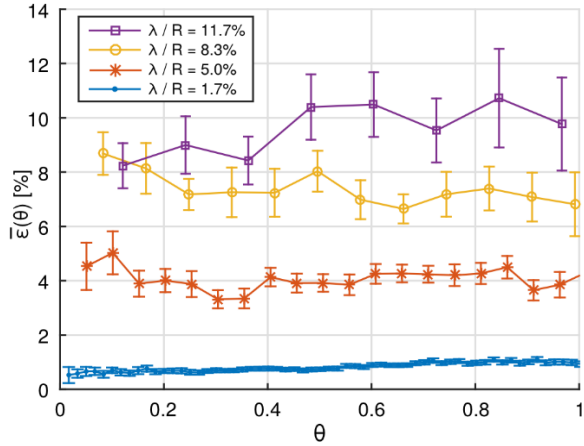

Figure 7. Error of the electric field magnitude along the surface with various mesh densities. Error bars show the standard error of $\bar{\epsilon}$ value within 95% confidence level.

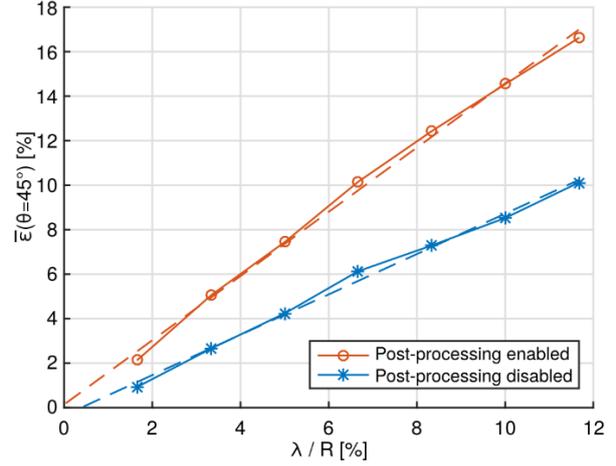

Figure 8. Error of the electric field magnitude near the surface with and without the field post-processor. Solid and dashed lines show the data points and their linear fitting, respectively, and demonstrate the solution convergence.

To show the effect of the field post-processing on the accuracy of the results, we measured the error $\bar{\epsilon}$ at a fixed polar angle of $\theta = 45°$, both with and without the post-processing. Figure 8 demonstrates the results of this measurement. As can be seen, the post-processor tends to decrease the field magnitude. For example, if the characteristic distance between the NN nodes is 10% of the hemisphere radius, the post-processing decreases the field by 6.8% in relation to the unprocessed value. Fitting the dependency (straight dashed line in figure 8) gives a more accurate estimation – the error of field with the post-processor is about 1.6 times higher than without it. This tendency is caused by the smoothing algorithm that replaces the field on the surface with the weighted average fields on and above the surface, where the field has always slightly smaller magnitude. Increasing the mesh density brings the hexahedral nodes closer to the tetrahedral ones and in the limiting case of infinitely dense mesh the numerical solution converges to the analytical one.

### 3.1.3. Solution accuracy on atomistic surface

To demonstrate the solution accuracy on the atomistic surface, we calculate the mean error $\bar{\epsilon}$ for all the tetrahedral nodes that lie on the atomistic hemisphere and inside the cone with semi-vertex angle of 60°. We repeat this measurement for various coarsening factors with and without using the field post-processing. The results are shown in figure 9. The graph again confirms the general trend that a coarser mesh results in less accurate field. However, in atomistic systems this trend is not strictly monotonous and its extent depends on the system size. Small systems with a small hemisphere radius tolerate only slight coarsening before the numerical solution quickly deviates from the analytical one. Larger systems, on the other hand, tolerate rather heavy coarsening without losing much of their accuracy.

It is noteworthy that both small and large systems tend to show more accurate results if at least a slight coarsening is applied to the mesh. Moreover, in dense systems with large hemisphere radius the field post-processor significantly increases the solution accuracy, while in small and coarse systems the results are more accurate without the post-processing. This can be explained by the FEM's sensitivity against sharp corners in the mesh – coarsening makes the mesh smoother and reduces the amount of non-physical artifacts in the solution before its post-processing. Therefore, we conclude that moderate coarsening of atomistic systems does not only reduce the problem size (thus reducing the computational cost), but also increases the solution accuracy. Moreover, at a given geometry and mesh density, the solution accuracy in large systems always exceeds the accuracy in small ones.

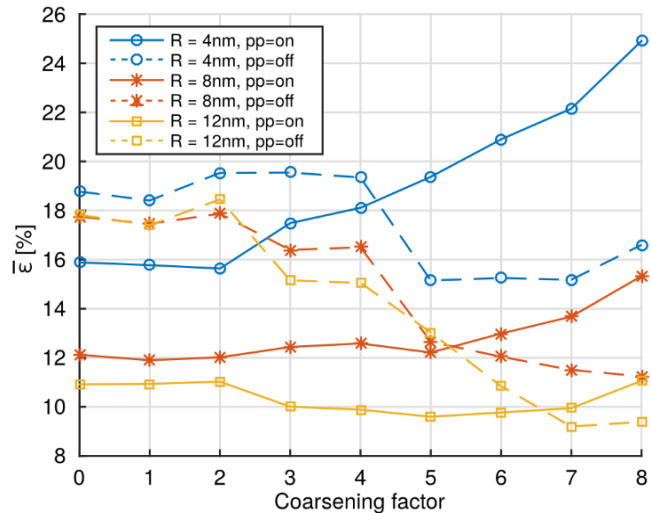

Figure 9. Error of the electric field near the atomistic surface with increasing coarseness and varying hemisphere radius *R*. Solid lines show the error if field post-processing is enabled and dashed lines depict the case where it is disabled.

### *3.2. Robustness of the results*

To demonstrate the robustness of our algorithms against the crystallographic structure of the material, we calculated the electric field around a nanotip with a molten apex. For this, we placed a cylindrical nanotip with a hemispherical cap on the substrate and ran a MD simulation to melt the nanotip apex. The system was cut out from a single-crystalline FCC block with ⟨100⟩ orientation and a lattice constant of 3.61 angstroms (Cu). The initial height and diameter of the tip was 24 nm and 6 nm respectively. Inside the tip, we applied a non-uniform ramp temperature distribution along the *z* direction with 600 K at the bottom and 1600 K at the top. As the simulation proceeded, the apex of the nanotip melted and formed an atomistic system with mixed amorphous-crystalline structure.

Figure 10 illustrates the three different stages of this simulation. Sections (a, b, c) show the evolution of the atomistic system, its local crystal structure as determined by common neighbor analysis [47] and the local electric field on the surface atoms. Sections (d, e, f) demonstrate the corresponding mesh that is generated around the nanotip. It is clearly visible that although a significant part of the nanotip is amorphous and has a transition region between amorphous and crystalline sections, our code manages to calculate a smooth electric field distribution in the whole simulation domain.

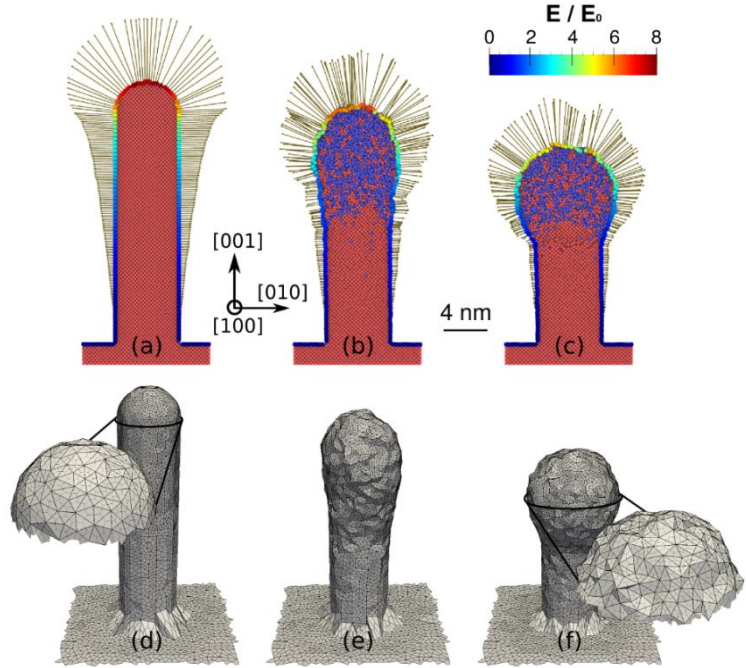

Figure 10. (a)-(c): cross-sections of the nanotip from different stages of the MD simulation. The surface atoms are colored according to the local electric field strength, while the color in the atomistic bulk region represents the local crystal structure as determined by common neighbor analysis; red – FCC, blue – amorphous. The arrows around the nanotip indicate the magnitude and direction of the local electric field. (d)-(f): the surface faces of the mesh at the same timesteps.

Finally, we ran a test to verify the robustness and stability of our method against small fluctuations in the input data. We created an FCC ⟨100⟩ surface with a single adatom on it. The adatom was placed in $n = 100$ random lattice positions near the center of the surface and the electric field was calculated for every case. Although the field distribution should not depend on the position of the adatom, it still fluctuates in the FEM calculation due to the rebuilding of the mesh for every iteration. The test showed that the field on the adatom where the mesh has the highest density of sharp corners fluctuates with a standard deviation of 1.7%.

### 3.3. Computational efficiency

For benchmarking purposes, we simulated atomistic hemispherical systems as described in section 3.1. We measured the variation of the code execution time for systems with different sizes and different coarsening factors (see figure 11 and figure 12). As all the linear dimensions of the system were chosen to be proportional to the radius $R$, the resulting number of atoms in the system scales roughly as $O(R^3)$. As can be seen, the computational cost is a sublinear function of the number of atoms in the system and decreases exponentially with increasing mesh coarsening level. Those results indicate that the computational efficiency (CPU time per atom) of our code increases as the system grows.

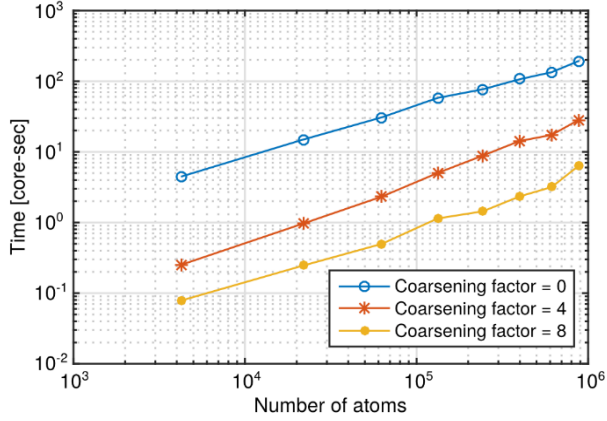 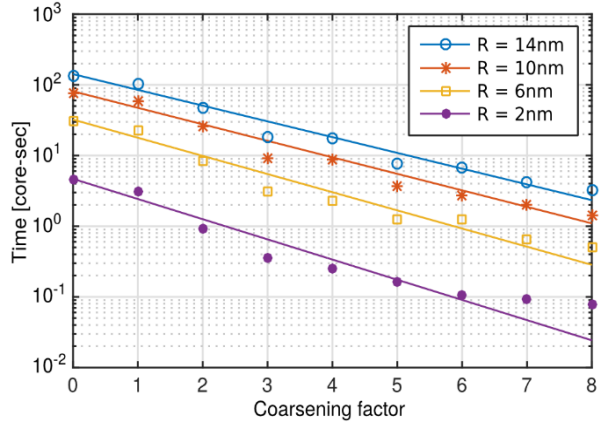

Figure 11. Code execution time dependence on the system size for various mesh coarsening factors.

Figure 12. Code execution time dependence on the mesh coarsening factor for systems with different size.

In section 2.8.2, we described a method for increasing the computational efficiency of a multiscale simulation. The core of this method is to measure the RMSD (14) for all atoms which have moved since the last full iteration and skip the next full iteration if the RMSD value is below the threshold. Here we demonstrate how this scheme affects the computation time and accuracy of the calculation.

For demonstration, we used the partially molten nanotip (see section 3.1.3), which has reached thermal equilibrium. At that stage, the shape of the nanotip was no longer changing significantly, while atomic motion was still relatively intense. We ran this system with different RMSD tolerances and measured the accuracy and computation time needed to simulate 4 ps of the nanotip evolution. In this test, we defined the accuracy as

$$\epsilon = \sqrt{\frac{1}{N} \sum_{i \in N} \left( \frac{E_i^j - E_i^{j-1}}{E_i^{j-1}} \right)^2}, \qquad (17)$$

where $E_i^j$ is the electric field norm in the location of *i*-th atom after the *j*-th full iteration and $N$ is the number of atoms in the system. To obtain statistics, the error (17) was averaged over all timesteps and plotted together with its standard deviation and total calculation time in figure 13. The graph reveals that the calculation time is inversely proportional to the RMSD threshold, while $\bar{\epsilon}$ increases proportionally with it. The best compromise between computation time and accuracy depends on the specific requirements of each simulation. However, a general way to obtain the optimal RMSD threshold is to pick one near the intersection point of the error and timing curves in figure 13.

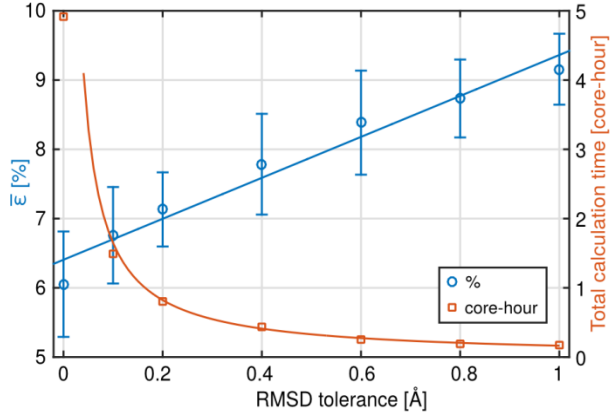
Figure 13. Solution accuracy and computation time as a function of atomistic RMSD threshold.

## 4. Discussion

The proposed code is designed to simulate various electronic processes near the surface of a nanostructure with any chemical composition and crystallographic orientation. The code outputs the electric field which affects the interatomic potential in MD and energy barriers in KMC simulations. To calculate those changes, the latter simulations require additional information on surface dipoles [48], while in MD, models similar to the ones suggested in [5], [6] can be used to approximate the charge induced on surface atoms. Furthermore, the high electric field initiates electron emission currents in the material, which may significantly affect the thermal evolution of the system [25], [29], [49] and cause electromigration and must therefore also be taken into account in atomistic simulations.

One important feature of the developed method is its ability to efficiently solve several differential equations of interest on the same mesh. The equations can be solved both in the vacuum and material domain and the solution can easily and efficiently be transferred between those regions. Thus, the method provides a framework for efficiently performing multiphysics calculations that are self-consistently coupled with large-scale atomistic simulations.

Currently we have implemented and verified only the 3D Laplace solver. However, preliminary tests have shown the possibility of also solving 3D heat and continuity equations, which would allow taking the effects of FE into account more accurately. Moreover, our method provides the framework for analyzing mechanical stress in simulations where this quantity is not inherent but still desired. For instance, the stress due to electric field can be introduced in KMC simulations.

## 5. Conclusions

We have developed a method to couple atomistic simulations with a finite element solver. Our algorithms dynamically build an unstructured mesh with optimized density that follows the material surface. After calculating the electric field and other physical quantities of interest on the mesh, the code exports the results back to the atomistic simulation. Our method provides the framework for efficiently, concurrently and self-consistently performing multiscale-multiphysics calculations.

**Acknowledgements**

The current study was supported by the Academy of Finland project AMELIS, Estonian Research Council Grants PUT 57 and PUT 1372 and the national scholarship program Kristjan Jaak, which is funded and managed by the Archimedes Foundation in collaboration with the Ministry of Education and Research of Estonia. We also acknowledge grants of computer capacity from the Finnish Grid and Cloud Infrastructure (persistent identifier urn:nbn:fi:research-infras-2016072533).

To speed up the development time, the current project takes advantage of several open-source and freely available software packages. The Delaunay tetrahedrization is performed with the C++ library Tetgen [36]. For FEM we use the C++ library Deal.II [41]. Several geometric operations are performed with the aid of C++ library CGAL [44]. The results of the simulations were visualized with the programs OVITO [50] and ParaView [51]. We thank all the people who have contributed to those projects.

**Appendix A. Summary of the code**

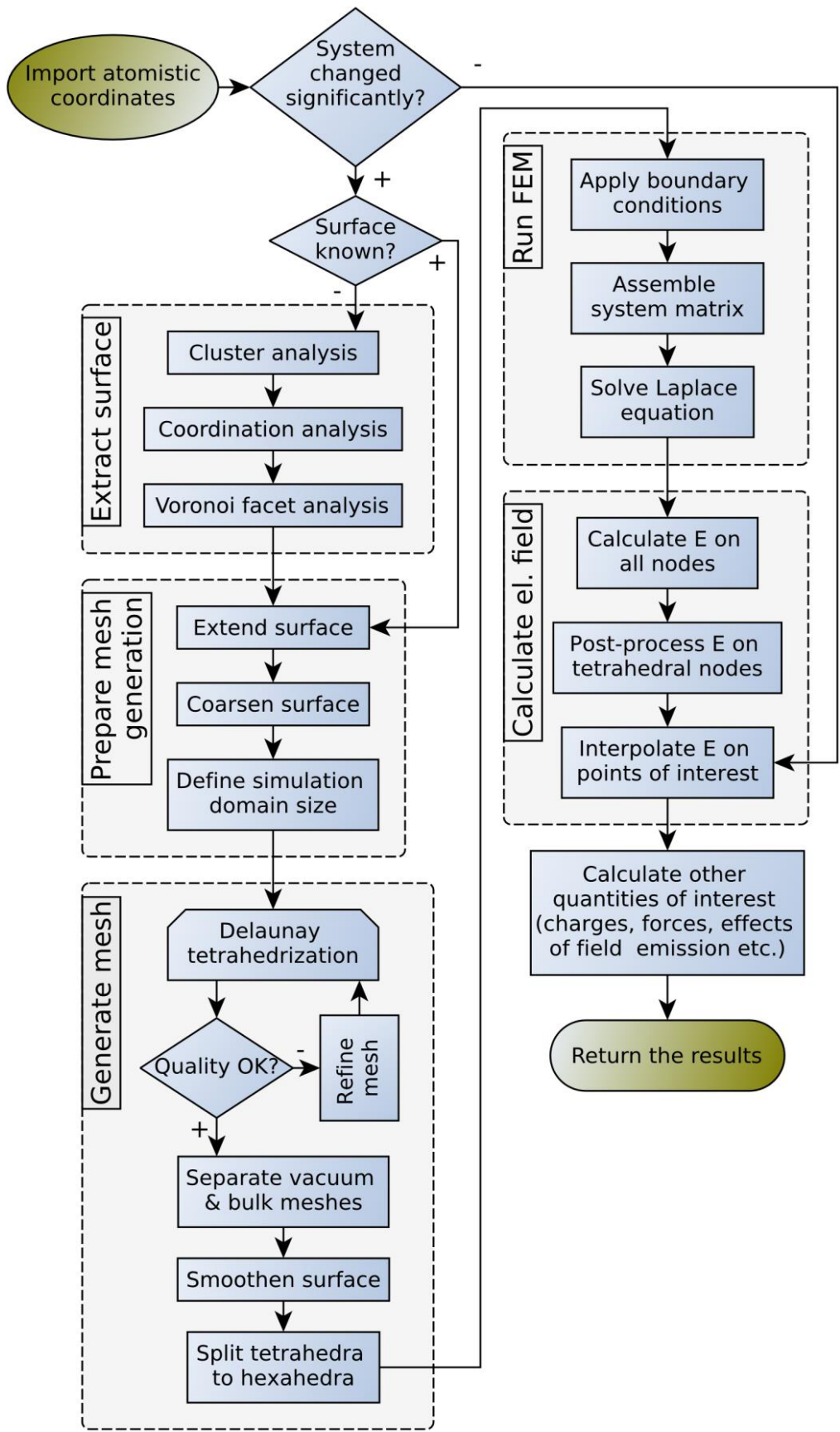

**Appendix B. Validating the results**

The electric potential Φ around a hemispherical protrusion residing on the center of the x-y plane can be expressed as [52]

$$\Phi(\boldsymbol{r}) = -E_0 \cdot z \cdot \left[1 - \left(\frac{R}{|\boldsymbol{r}|}\right)^3\right], \quad (18)$$

where $R$ is the radius of the hemisphere and $E_0$ is the long-range applied electric field. By plugging equation (18) into (5), we obtain the analytical distribution of the electric field:

$$\frac{\boldsymbol{E}_a(\boldsymbol{r})}{E_0} = \frac{3R^3 xz}{|\boldsymbol{r}|^5}\hat{\boldsymbol{x}} + \frac{3R^3 yz}{|\boldsymbol{r}|^5}\hat{\boldsymbol{y}} + \left(1 - R^3 \frac{x^2 + y^2 - 2z^2}{|\boldsymbol{r}|^5}\right)\hat{\boldsymbol{z}}. \quad (19)$$

Denoting the local field near the nanostructure apex ($x = y = 0$, $z = R$) as $E_{apex}$, we define the field enhancement factor as

$$\gamma = \frac{E_{apex}}{E_0}. \quad (20)$$

According to equation (19), the analytical value of $\gamma$ factor for a hemisphere on a planar surface is

$$\gamma_a = 3. \quad (21)$$

The numerically calculated field values depend on the anode-cathode distance $d$ and the half-width $D$ of the substrate (see figure 5). Equation (21) is valid if $d$ and $D$ are infinite. In simulations, however, finite $d$ and $D$ must be used, thus causing systematic error in the results. We can estimate this systematic error by comparing long-range field values to the ones on the system boundaries, where constraints have been imposed. Doing so, implies

$$\frac{E_a(0,0,d)}{E_a(0,0,\infty)} = 1 + 2\left(\frac{R}{d}\right)^3, \quad (22)$$

$$\frac{E_a(D,0,0)}{E_a(\infty,0,0)} = \frac{E_a(0,D,0)}{E_a(0,\infty,0)} = 1 - \left(\frac{R}{D}\right)^3. \quad (23)$$

Given the stability of the Laplace equation, the above equations give a rough estimation of how the overall error depends on the system size. Therefore, to keep the estimated systematic error in our simulations below 1%, we always apply the long-range electric field at least 6 times higher than the nanostructure height ($d \geq 6R$) and make the simulation box at least 5 times wider ($D \geq 5R$).